\documentclass[]{interact}

\usepackage[natbibapa,nodoi]{apacite}
\theoremstyle{plain}

\theoremstyle{definition}

\theoremstyle{remark}

\graphicspath{{figures/}}

\usepackage{hyperref}
\usepackage{subcaption}

\usepackage{amsmath}
\usepackage{mathtools}

\newcommand{\cparagraph}[1]{\par\vspace{2mm}\noindent\textbf{#1}}

\begin{document}

% \articletype{ARTICLE TEMPLATE}

\title{Summarizing Classed Region Maps with a Disk Choreme}

\author{
 \name{Steven van den Broek\,\textsuperscript{a}\thanks{CONTACT Steven van den Broek. Email: s.w.v.d.broek@tue.nl}, Wouter Meulemans\,\textsuperscript{a}, Andreas Reimer\,\textsuperscript{a,b}, Bettina Speckmann\,\textsuperscript{a}}
 \affil{
  \textsuperscript{a }TU Eindhoven, Netherlands\\
  \textsuperscript{b }Arnold-Bode-Schule Kassel, Germany
}
}

\maketitle

\begin{abstract}
Chorematic diagrams are highly reduced schematic maps of geospatial data and processes. They can visually summarize complex situations using only a few simple shapes (choremes) placed upon a simplified base map. 
Due to the extreme reduction of data in chorematic diagrams, they tend to be produced manually; few automated solutions exist.
In this paper we consider the algorithmic problem of summarizing classed region maps, such as choropleth or land use maps, using a chorematic diagram with a single disk choreme.
It is infeasible to solve this problem exactly for large maps.
Hence, we propose several point sampling strategies and use algorithms for classed point sets to efficiently find the best disk that represents one of the classes.
We implemented our algorithm and experimentally compared sampling strategies and densities. The results show that with the right sampling strategy, high-quality results can be obtained already with moderately sized point sets and within seconds of computation time.
\end{abstract}


\section{Introduction}
\label{sec:introduction}

Thematic maps are the tool of choice to inspect, analyze, and communicate spatial data. In such maps, full geographic accuracy (in so far that it exists) can distract from or even obscure higher-level patterns. In fact, full detail is not necessary or even desirable in many settings. \emph{Chorematic diagrams} offer highly schematized representations of geographic data and processes. Originally introduced by \cite{brunet1980}, they consist of a simplified or schematic base map and one or more layers of data visualizations, using fixed symbolism (\emph{choremes}) to capture the most useful or salient aspects \citep{reimer2010understanding}. Chorematic diagrams are used, for example, to visually summarize detailed maps as insets, they accompany essays explaining geographic phenomena, and they can act as a preview thumbnail in a database of maps. We refer the reader to \cite{andreasthesis} for a more extensive exposition on chorematic diagrams.

\begin{figure*}
    \includegraphics[width=0.376\linewidth]{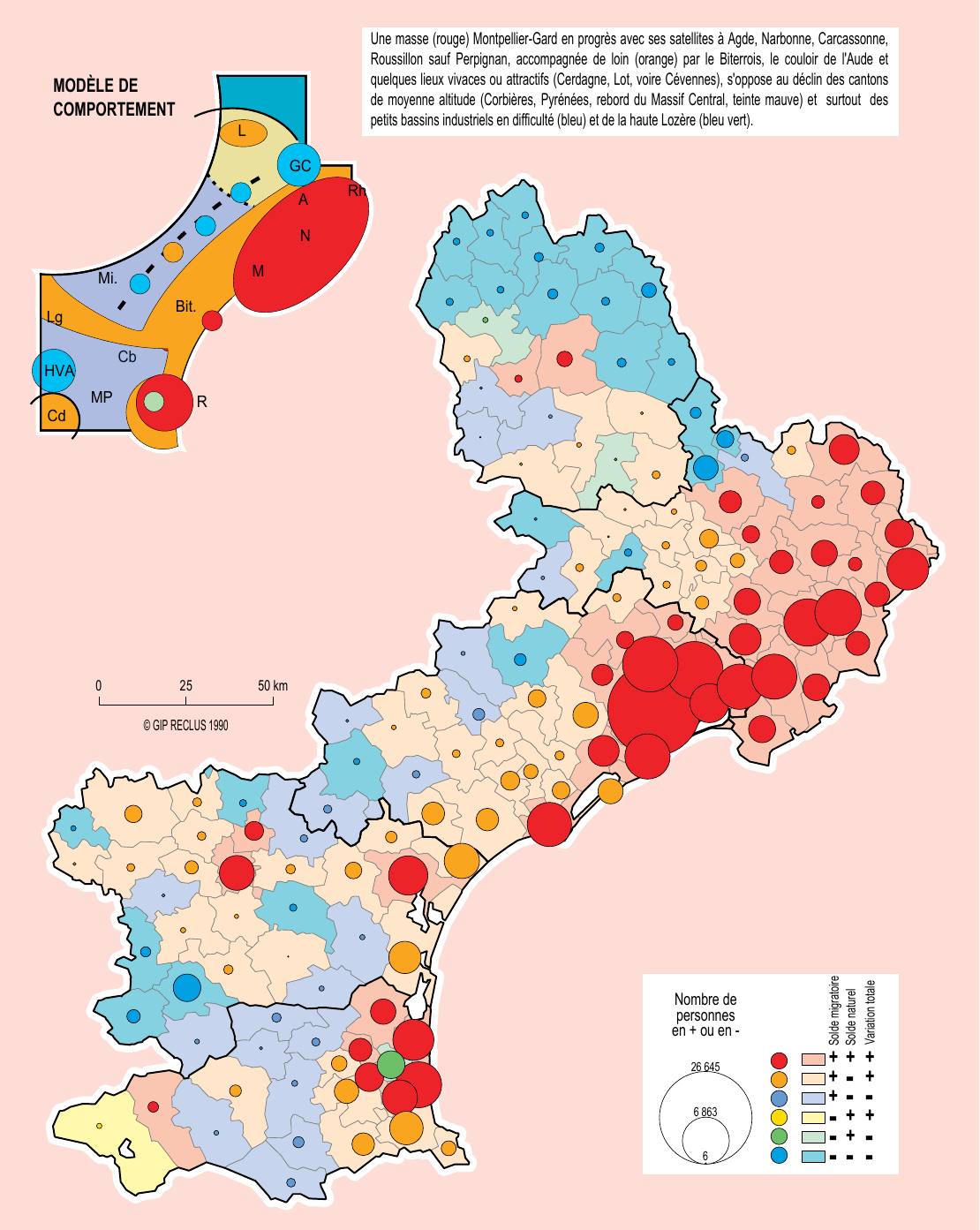}
    \hfill
    \includegraphics[width=0.59\linewidth]{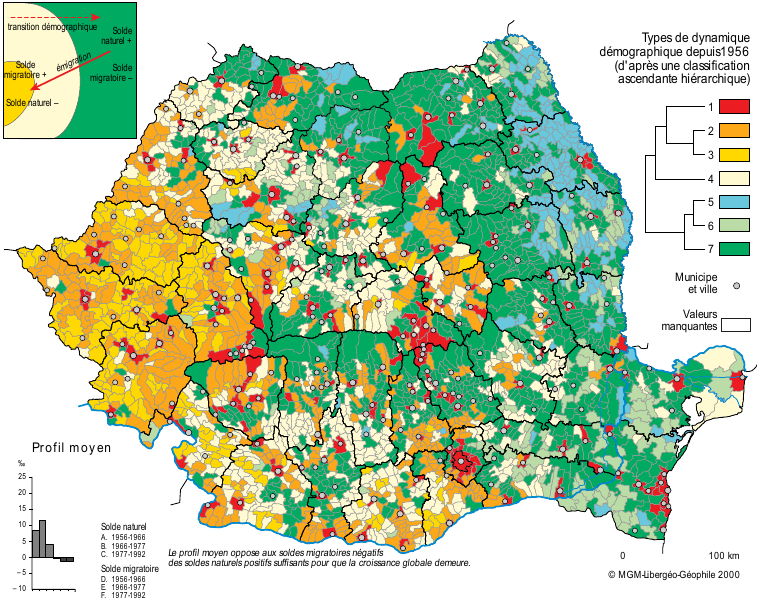}
    \caption{Example chorematic diagrams summarizing a choropleth map. Left: Languedoc-Roussilon in France \citep{Brunet1991}. Right: Romania \citep{rey2000}, cropped from \cite{andreasthesis}.}
    \label{fig:LR-example}
\end{figure*}

Chorematic diagrams are traditionally constructed by hand, but doing so is time-intensive and precludes their use in interactive visual analytics systems that require instantaneous response to user queries. While automated solutions are clearly desirable, they are currently mostly lacking due to the inherent challenge to design algorithms that can capture complex processes in very simple shapes.
The work by \cite{DelFatto2009} and \cite{DeChiara2011} is a first step towards 
interactive chorematic diagrams: they design a system for representing, creating, and interacting with choremes. 
However, algorithms for automated construction of high-quality chorematic diagrams are still missing.
There has been previous work that targets another aspect of chorematic diagrams, namely the very reduced representation of regions \citep{parallellism,stenomaps}.
Chorematic diagrams are an extreme form of cartographic schematization; we treat related work in this area more extensively below.

\begin{figure}[b]
    \centering
    \includegraphics[page=2]{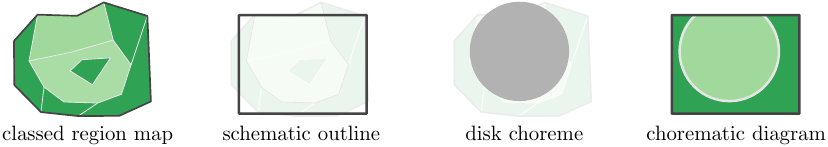}
    \caption{A chorematic diagram consists of two components: a schematic outline and a set of choremes. We focus on representing a single class of a classed region map using a single disk choreme.}
    \label{fig:terms}
\end{figure}

\cparagraph{Contributions.} 
Inspired by \autoref{fig:LR-example}, we take a first step towards automatically computing high-quality chorematic diagrams. We focus on classed region maps, in which each region is assigned a class based on data, such as  choropleth maps, land use maps, and area-class maps. Specifically, we study how to represent a single class of a classed region map using a single symbol: a disk choreme (\autoref{fig:terms}).
We first model the algorithmic problem in \autoref{sec:problem}. Exact solutions for maps with polygonal regions are computationally expensive and do not scale well. Hence we present a sampling approach which approximates classed regions with classed sets of points. In \autoref{sec:sampling} we describe our algorithm, including various sampling strategies. \autoref{sec:experiments} evaluates the efficacy of these strategies, and shows that our method can achieve high-quality results already with moderately sized point sets within second of computation time.
We close in \autoref{sec:conclusion} with future work.

\cparagraph{Related work.} 
As mentioned above, computing chorematic diagrams is an extreme form of schematization. The literature on cartographic schematization is extensive; here we restrict ourselves to mentioning some examples of methods that compute schematic outlines, which can then form the base map for chorematic diagrams \citep{parallellism,goethem2015exploringcurved,bmrs2016}.

Finding a single disk choreme that summarizes one class well is a form of shape matching: transforming one shape to maximize its similarity to another (scaling and translating a disk so that it matches a class as best as possible).
Also shape matching is studied extensively and many approaches exist that vary according to the similarity measure used and the transformations supported, see \citep{DBLP:journals/amai/AltBB95} for a representative example. Shape matching is a challenging problem and hence algorithmic solutions are often complex or restrict themselves to simpler variants such as convex shapes \citep{DBLP:conf/icalp/AltBW90, yon2016approximating}.
We are hence exploring approximate solutions via point sampling. Here we need to solve the following two problems: $(1)$ represent a set of polygons of a particular class by points, $(2)$ (re)construct a (disk) shape from a point set.

Quantitative methods in land surveying, geography, and other spatial sciences \citep{bunge,haggett} collect point data for the reconstruction of scalar fields and other phenomena within polygonal regions. Here sampling generally takes place according to a lattice; the structure of this lattice and the parameters that govern it necessarily depend on the phenomena to be captured. See \cite{delmelle} for a current short overview  or \cite{brus} for an applied, in-depth introduction. 

Reconstructing a shape from point samples is another well-studied problem. Methods are based either solely on the points within the shape, e.g.\ \citep{edelsbrunner1983shape,duckham2008efficient,meulemans2011imprecise}, or also on points to be excluded, e.g.\ \citep{fisk1986separating,edelsbrunner1988minimum,reinbacher2008delineating,
DBLP:conf/iccsa/BeregDZR15}.

A final note: classed region maps, such as choropleth maps, suffer from the problem that feature size does not necessarily represent relevance or data magnitude; this is known as the modifiable area unit problem \citep{maup}. Single disk chorematic diagrams inherit this issue; despite the extreme simplification they do not deform the underlying space as, for example, cartograms do.

\section{Chorematic diagrams for classed region maps}
\label{sec:problem}

In this section, we formalize the notion of creating a chorematic diagram for a classed region map in an algorithmic problem statement. 
Our input is the classed region map, which we represent as a set $\mathcal{R}$ of polygonal regions. 
Each region is a tuple $(P,c)$ of its polygonal shape $P$ (possibly with holes and multiple components) and its class $c$. 
We assume that the polygons are pairwise interior-disjoint.

To summarize a classed region map in a chorematic diagram, one may want to use multiple of a variety of elementary shapes (choremes) such as disks, ellipses, or annuli. 
In this paper we focus on the basic problem of placing a single disk choreme to visually summarize a single class. We simplify the discussion by considering a region map with only two classes; at the end of this section we discuss how our approach generalizes to more classes.
Our goal is to capture the visual structure of the classed region map; thus, we concern ourselves only with the classes, and not with the underlying data that gave rise to them.

\begin{figure*}[t]
    \centering
    \includegraphics{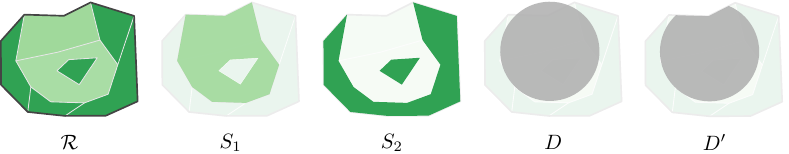}
    \caption{Illustrations of notation on an example region map with two classes. Set $\mathcal{R}$ consists of polygonal regions; $S_1$ and $S_2$ are sets of all points of class $1$ (light green) and $2$ (dark green) respectively. $D$ is a disk; $D'$ is the subset of $D$ that lies within the map $\mathcal{R}$.}
    \label{fig:notation}
\end{figure*}

Let $c_1$ and $c_2$ be the two classes, and let $S_1$ and $S_2$ be the sets of points in the map $\mathcal{R}$ that belong the classes $c_1$ and $c_2$ respectively. That is, $S_i = \bigcup_{(P, c_i) \in \mathcal{R}} P$. 
See \autoref{fig:notation} for illustrations of these and upcoming definitions.
Our goal is to represent $c_1$ with a disk~$D$ that overlaps $S_1$ as much as possible. That is, we want to maximize $|S_1 \cap D|$, which denotes the area of $S_1$ within disk $D$.
At the same time, we want to avoid suggesting class $c_1$ for regions that are overlapped by $D$ but belong to $c_2$. That is, we want to minimize $|S_2 \cap D|$.
We assume that the disk is to be constrained to the map region $\mathcal{R}$ eventually. As such, we disregard overlap between $D$ and area outside of any regions.

As both $|S_1 \cap D|$ and $|S_2 \cap D|$ measure area, we combine them and aim to maximize, for some choice of $\alpha \in [0, 1]$:
\[\alpha \cdot |S_1 \cap D| - (1 - \alpha) \cdot |S_2 \cap D|.\]
We use \emph{score} to refer to the value of this objective function.
If $\alpha = 0.5$ then the score gained by covering $c_1$ is equal to the penalty of covering $c_2$. In this case, the objective function is equivalent to the symmetric difference, a concept from set theory.
Indeed, let $D'$ denote the set of points in disk $D$ that lie within the map $\mathcal{R}$.
Then, because any part of $D'$ that is not part of $S_1$ is part of $S_2$, the formula can be rewritten to
\begin{align*}
   &\alpha \cdot (|S_1| - |S_1 \setminus D'|) - (1-\alpha) \cdot |D' \setminus S_1|\\
=\ &\alpha \cdot |S_1| - (\alpha \cdot |(D' \setminus S_1)| + (1-\alpha) \cdot |(S_1 \setminus D')|).
\end{align*}
Thus, when $\alpha = 0.5$ maximizing the objective function is equivalent to minimizing the symmetric difference between $D'$ and $S_1$.

\begin{figure}[t]
    \hspace*{\fill}
    % \hspace{10mm}
    \begin{subfigure}{0.2\textwidth}
        \includegraphics[width=\textwidth, page=3]{trypo.pdf}
        
        \vspace{2mm}
        
        \includegraphics[width=\textwidth, page=2]{trypo.pdf}
    \caption{Synthetic}
    \end{subfigure}
    \hfill
    \begin{subfigure}{0.7\textwidth}
        \includegraphics[width=0.49\textwidth, page=1]{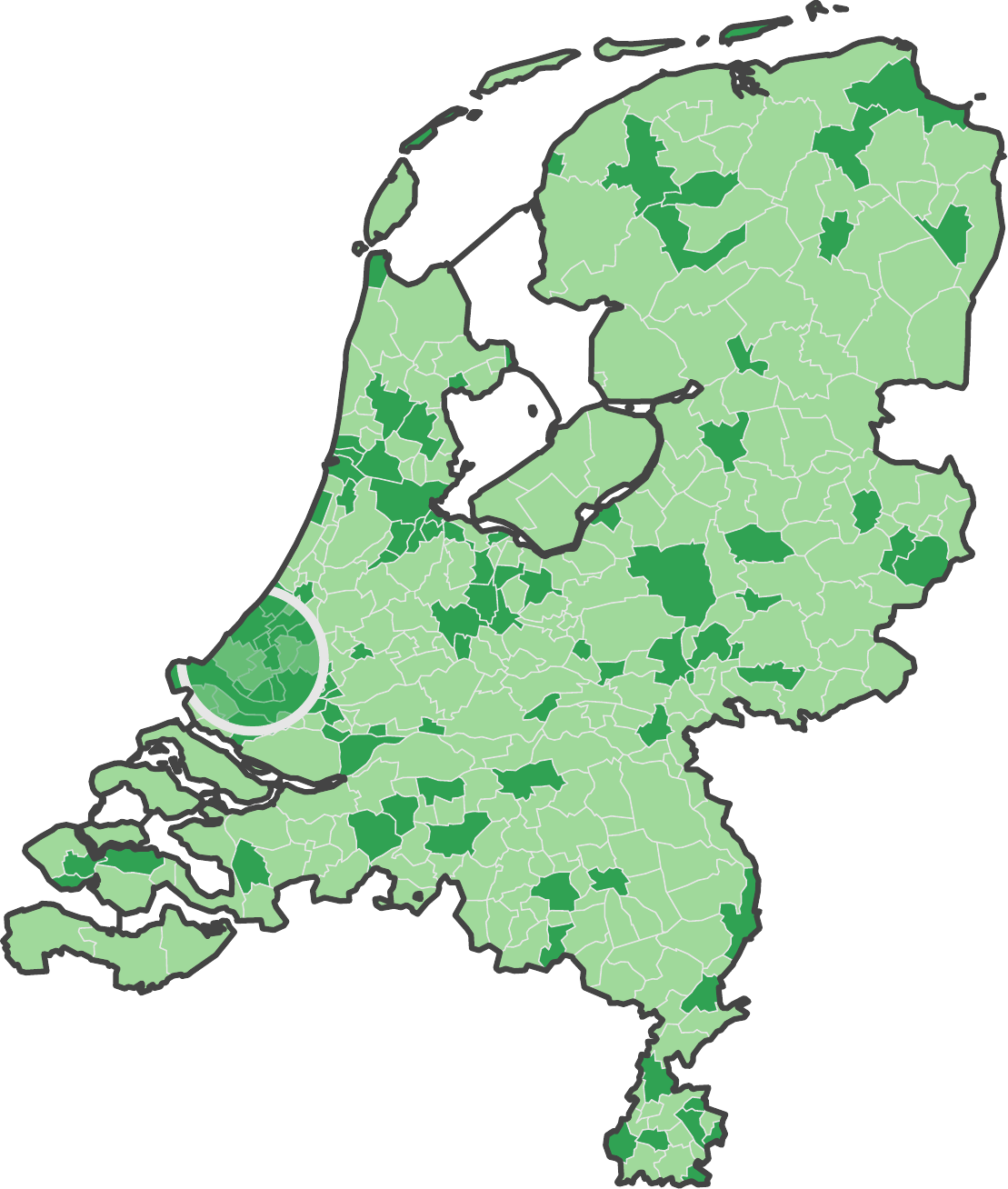}
        \includegraphics[width=0.49\textwidth, page=2]{percentage_huurwoningen_voronoi_25_1000.pdf}
        \caption{Real-world}
    \end{subfigure}
    \hspace*{\fill}
    \caption{Two examples that highlight the difference between setting $\alpha = 0.5$ and using our proposed weights. Top/left: for $\alpha=0.5$, the optimal disk describing the dark green regions captures only a small fraction. Bottom/right: using our proposed weights, the optimal disk captures a larger portion of the dark green regions, at the expense of covering more of the (more frequent) lighter green areas.}
    \label{fig:weights}
\end{figure}

When $\alpha = 0.5$, the two classes are treated symmetrically.
Consequently, when there is relatively little of $S_1$ and it is spread across the map, then an optimal disk $D$ will be small to avoid overlap with $S_2$.
See \autoref{fig:weights} for an illustration on two example maps, one synthetic and one from real-world data.
The optimal disk according to the symmetric difference is rather arbitrary and at best identifies a core area of the map where class $c_1$ is most prominent.
As our intent is to let the optimal disk summarize the entire class~$c_1$, we use a different value for scalar $\alpha$ such that more of $c_1$ will be covered by the disk. 
Intuitively, we set the scalar $\alpha$ such that the penalty and gain for covering the regions depends on the rarity of the corresponding class.
We set $\alpha = |S_2| / (|S_1| + |S_2|)$.
Thus, there is an inverse relationship between the effect of covering a class and its presence on the map.
Our choice for $\alpha$ in a sense negates any imbalance in the proportion of area covered by the classes; note that, as a result, a disk that covers the entire map has score zero.

\cparagraph{General form.} 
In general, we may consider that regions have a value from some set $\mathcal{V}$, and that there is a distance measure $\delta$ on $\mathcal{V}$ that indicates the similarity between two values in $\mathcal{V}$. We assume that $\delta$ gives positive values for regions that are to be covered by the disk, and negative for regions that are not to be covered by the disk. We then obtain the following general form for the quality of a disk $D$ with value $v$:
\[ \sum_{(P,v') \in \mathcal{R}} \delta(v,v') |P \cap D|. \]
For our two-class region map, we have hence used $\delta(v,v') = \alpha$ if $v = v'$ and $\delta(v,v') = \alpha - 1$ otherwise. 
For region maps with more classes, one could use the distance measure $\delta$ to capture how dissimilar two classes are, similar to area aggregation \citep{haunert2007optimization, gedicke2021aggregating}.

\section{Approximate solutions via sampling}
\label{sec:sampling}
\begin{figure*}[tb]
    \centering
    \includegraphics[page=1]{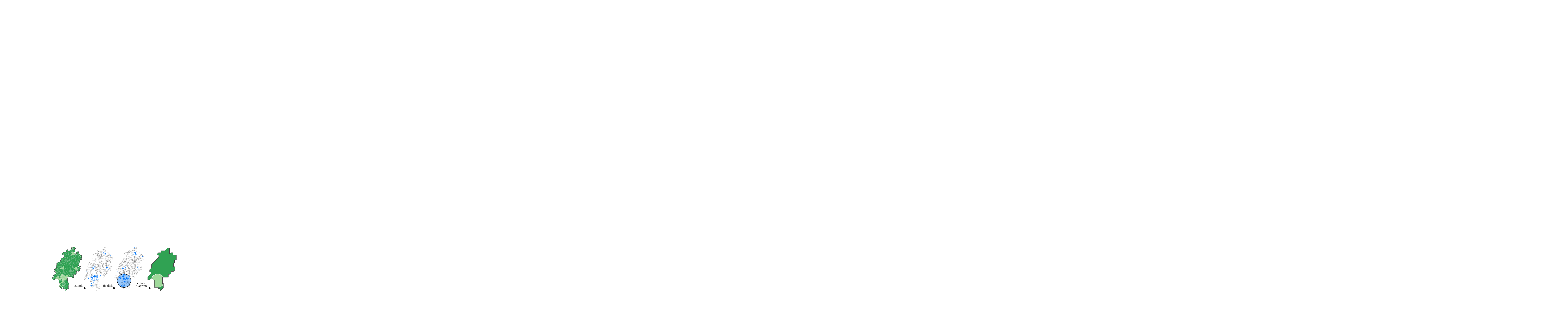}
    \caption{Pipeline for summarizing a classed region map with a disk. First, we sample points on the map. We assign a weight to each point based on the region it lies in and the class we are summarizing. In the figure, positive weight points are drawn as blue disks, and negative weight points as grey crosses. We find the smallest maximum-weight disk on this weighted point set (supported by at least two blue points) which serves as an approximation for the optimal disk on the region map.
    The figure shows $1000$ points sampled in Hesse using the local Voronoi approach with 25 iterations.
    }
    \label{fig:pipeline}
\end{figure*}

Solving the problem precisely for polygonal regions is cumbersome, due to, among other aspects, the lack of an analytic solution to the involved equations \citep{kreveld2004diagramsonmaps}. Hence, we take a two-step approach: (1) we sample map $\mathcal{R}$ to obtain a point-based approximate representation; (2) we solve the problem exactly using the point representation. We explain these two steps in the following subsections; see also \autoref{fig:pipeline}.

\cparagraph{Related theory.} 
In the field of computational geometry there is theoretical work that provides guarantees regarding the use of point samples to approximately solve geometric problems that deal with intersections, such as the one we are interested in solving in this paper. 
As long as the type of intersection is not too complicated, formalized by the Vapnik-Chervonenkis dimension (shortened as VC-dimension), solving a problem on points sampled in the universe (the map in our case) uniformly at random provides an approximation, and there are bounds on the size sample one needs to attain a certain level of approximation.
Though such work, in particular $\varepsilon$-approximations (for an overview, see \cite{mustafa2017epsilon}), is applicable, their guarantees hold only for very large sample sizes.
As we shall see, our second step still involves a nontrivial algorithm and thus we are particularly interested in strategies for generating small to medium-sized point sets that achieve high quality in the second step.  

\subsection{Step 1: Sampling a map}

Below, we describe our sampling strategies. Each method can be applied either \emph{globally} to the entire choropleth map, or \emph{locally} to (components of) regions separately.

With a local strategy, the number of points we sample in a component is directly proportional to its area.
Our approach to determining the number of points to sample in each component is as follows. 
Let $n$ be the total number of points we want to sample, let $C_1, \dots, C_k$ denote the components, and let $A$ be the total area of the map.
We create a `bin' $n_i$ for each component $C_i$, which is an integer that represents the number of points to be sampled in $C_i$.
We divide integer $n$ over the bins such that $\sum_i n_i = n$.
If sample points were divisible into fractional pieces then the proportion of points to sample in $C_i$ would simply be equal to the proportion of the map covered by $C_i$; that is, $n_i$ would be $p_i \coloneq |C_i| / A \cdot n$. 
As they are not, we first add to each bin $n_i$ the number of whole points $\lfloor p_i \rfloor$ that fit in $p_i$.
We then sort the bins $n_i$ on descending order of the fractional part of $p_i$, and distribute the remaining points one by one in that order. 
This approach minimizes the average and maximum difference between $n_i$ and $p_i$.

\begin{figure}[t]
    \begin{subfigure}{0.24\columnwidth}
        \centering
        \includegraphics[page=1]{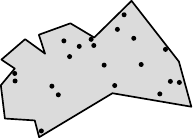}
        \caption{Random}
    \end{subfigure}
    \hfill
    \begin{subfigure}{0.24\columnwidth}
        \centering
        \includegraphics[page=2]{sampling.pdf}
        \caption{Voronoi}
    \end{subfigure}
    \hfill
    \begin{subfigure}{0.24\columnwidth}
        \centering
        \includegraphics[page=3]{sampling.pdf}
        \caption{Square grid}
    \end{subfigure}
    \hfill
    \begin{subfigure}{0.24\columnwidth}
        \centering
        \includegraphics[page=4]{sampling.pdf}
        \caption{Hexagonal grid}
    \end{subfigure}
    
    \caption{Sampling strategies.}
    \label{fig:sampling-strategies}
\end{figure}

We now turn to describing our four sampling strategies.
In our description below, we assume that the shape to be sampled is a polygon $P$.
In a global implementation, $P$ is the union of all regions, which may have holes and multiple components; in a local implementation, $P$ is a region component.
\autoref{fig:sampling-strategies} illustrates our sampling strategies.

\cparagraph{Random.} 
With this approach, we sample a given number of points from $P$, uniformly at random. We do so by triangulating $P$, choosing a triangle randomly weighted proportionally by its area, and then choosing a point uniformly within the triangle. Using this implementation, each point in $P$ is equally likely to be selected.

\cparagraph{Voronoi.}
With this approach, we first generate a sample using the random method. We then postprocess this sample to improve how well spread the points are.
We use the well-known algorithm by~\cite{DBLP:journals/tit/Lloyd82} to iteratively move the points.
We do not allow sample points to move to a different component of $P$; therefore, we execute this algorithm per component separately, in parallel.
Specifically, we repeat the following steps for a component $C$ of $P$ for a fixed number of iterations:
\begin{enumerate}
    \item Compute the Voronoi diagram of the current set of points.
    \item Crop the Voronoi diagram by intersecting it with $C$. 
    \item Move each point to the centroid of its cropped Voronoi cell.
\end{enumerate}
This sampling approach approximates a centroidal Voronoi tesselation~\citep{DBLP:journals/siamnum/DuEJ06}.
Note that we use a Voronoi diagram of points intersected with $C$ as an approximation of the geodesic Voronoi diagram of points within $C$ \citep{DBLP:journals/algorithmica/Aronov89}.
We chose this as software implementations of standard Voronoi diagrams are readily available and they work well for the purposes of spreading out sample points.
This approach is related to the spatial coverage sampling technique used by \cite{brus2006designing}; the difference is that they discretize the polygon $C$ before iteratively moving points.

\cparagraph{Square grid.}
Given a grid size $s$, we generate a regular grid of squares with side length $s$ within the bounding box of $P$.
We align the bottom left of the square grid with the bottom left of the bounding box of $P$.
We add to our sample each center of a square that lies within $P$.

Note that we do not directly control the number of points created for a particular region, only indirectly through the grid size $s$. Intuitively, the grid size is inversely proportional to the number of obtained samples.
However, this relation is not perfectly so: slightly increasing grid size may lead to more samples, just due to some centers shifting into the polygon.
Nonetheless, we can attempt a binary search on $s$ to obtain a given number of points. In our experiments, we always obtained the exact result. In actual application, the exact number may be less relevant, as long as the obtained quality of the disk is high.

Our alignment of the grid may create minor bias to the bottom-left, compared to, for example, centering the grid at the center of the bounding box.
However, the alignment we use results in a continuous movement of the points as the grid size increases, which aids the binary search in finding a grid size for a target number of samples.

\cparagraph{Hexagonal grid.}
This method is the same as the above, except that we use the centers of a regular hexagonal grid, using hexagons of side length $s$ and with a vertical side.

\cparagraph{Rationale.}
The grid methods give a good spread of points for simple shapes. However, they are prone to miss large parts of a polygon if its boundary is highly irregular. That is, a part of a polygon of large area may receive comparatively few samples, if the grid points just happen to lie outside it.
The uniform method avoids this issue by ensuring that any point in a polygon has equal probability of being chosen. However, its random nature may cause the points to not be spread out as well throughout a polygon, even for simple shapes.
The Voronoi method avoids both issues by its iterative process of spreading points throughout the polygon. 
As such, we expect it to give good results at the cost of a significantly higher running time than other methods. 

Global strategies may be simpler and readily encapsulate that regions of different sizes receive different number of samples. However, they are also more prone to the randomization issues mentioned above, and are more time-consuming to compute. As such, we expect local strategies to be more effective overall.

\subsection{Step 2: Computing with a point set}
In \autoref{sec:problem} we formulated a maximization problem for fitting a disk to a class $c$ of a two-classed region map.
One can view the problem as acting on a set of weighted regions: a region of class $c$ has weight $\alpha$, and other regions have weight $\alpha - 1$.
To approximate this problem using points we similarly create a weighted set of points.
We assign to a sample point $p$ the weight of the region $p$ lies in.

\cparagraph{Using weighted points.}
We now have a weighted point set $\mathcal{P}$, where each weighted point is a tuple $(p,w)$ of a coordinate $p \in \mathbb{R}^2$ and weight $w \in \mathbb{R}$. Our purpose is to place the disk such that we accrue as much weight as possible: positive weight implies we want the disk to cover the point, negative-weight points should be avoided.
So, our goal is compute a disk $D$ such that $\sum_{(p,w) \in \mathcal{P} \cap D} w$ is maximized. 

This problem is effectively solved by \cite{DBLP:conf/iccsa/BeregDZR15}. 
As there are an infinite number of optimal disks they focus on finding the smallest one.
The idea is that the smallest maximal disk must have two positive-weight points on its boundary. Pick any such pair: the center of $D$ must lie on their bisector. We can sweep the center along this bisector. The weight of the disk during this sweep changes when points enter or exit the disk. We execute this sweep using the appropriate events while keeping track of the total weight in the disk. This takes $O(n \log n)$ time for one pair, and thus $O(n^3 \log n)$ time for trying all pairs, where $n$ is the total number of points in $\mathcal{P}$.

We observe that this algorithm can easily be parallelized, as each of the sweeps is independent. In our experiments (\autoref{sec:experiments}), we use a simple parallelized implementation.

\section{Experimental evaluation}
\label{sec:experiments}
We implemented the sampling strategies for experimental evaluation. Our implementation builds on CartoCrow\footnote{\url{https://algo.win.tue.nl/software/cartocrow/}} and CGAL for robust geometric computations~\citep{cgal:wfzh-a2-24b, cgal:y-t2-24b}; the program is publicly available\footnote{\url{https://github.com/tue-alga/cartocrow}}.

\cparagraph{Methodology.} After a brief consideration of the running time for fitting a point to a weighted set of points, the bottleneck in our computations, we focus on the differences in performance between our sampling strategies in terms of quality difference. 

To this end, we run each of the sampling strategies on the collected datasets (see below), using sampling numbers ranging from $100$ to $1000$, and compute the smallest maximum-weight disk of each sample. We then establish the quality of the disk using the full polygonal shapes.

To obtain an aggregate view of the performance of the various methodologies, we define the \emph{relative quality} of a disk as the ratio between its score and the best score obtained over all strategies and sampling numbers. The best score obtained serves as a proxy for the actual optimal result. To ensure that this is a high-quality proxy, we also include the results of the Voronoi method with $10\,000$ points.

We are interested in understanding how much the relative quality improves as the number of samples increases, but specifically also to uncover which strategy performs well most consistently. 

\cparagraph{Data.} 
To evaluate our sampling methods, we use twelve choropleths. Specifically, we obtained two sets of administrative boundaries with associated statistical data: the 345 Dutch municipalities in 2022\footnote{\url{https://www.cbs.nl/nl-nl/dossier/nederland-regionaal/geografische-data/wijk-en-buurtkaart-2022}; © Kadaster / Centraal Bureau voor de Statistiek, 2024.} and the 425 municipalities in the state of Hesse, Germany\footnote{\url{https://daten.gdz.bkg.bund.de/produkte/vg/vg250-ew_ebenen_1231/aktuell/}; © Bundesamt für Kartographie und Geodäsiem 2024, under \href{https://www.govdata.de/dl-de/by-2-0}{dl-de/by-2-0}.}. From the statistical data\footnote{\url{https://statistik.hessen.de/}; © Hessisches Statistisches Landesamt, Wiesbaden.}, we selected six attributes for the Netherlands and six for Hesse. 
Each of the attributes was split into two categories using the natural breaks method~\citep{fisher1958grouping, jenks1967data} which minimizes intra-class variance and maximizes inter-class variance.
The attributes were selected arbitrarily, but such that the resulting patterns were visually distinct from one another.

The geometry of the maps were generalized to have $5000$ edges for the Netherlands and $3824$ edges for Hesse.
To facilitate reuse and reproducibility, we have made the data on which we ran our experiments publicly available\footnote{\url{https://doi.org/10.5281/zenodo.15524996}}.

\cparagraph{Running time.}
Our technique, regardless of sampling strategy, needs to fit a disk to the obtained weighted point set. As the algorithm takes cubic time, its practical scalability in terms of running time needs to be investigated. To this end, we generated random point sets of $1000$ points, half of which are positive and half of which are negative. The sequential implementation takes approximately\footnote{On a laptop with 32GB of RAM and with as CPU an 11th Gen Intel(R) Core(TM) i7-11800H @ 2.30GHz processor (8 cores).} 12.6 seconds, whereas the parallel implementation takes 1.4 seconds. These running times are certainly feasible in semi-interactive systems. The parallel implementation takes for $2000$ points and $10\,000$ points 13.3 seconds and 26.3 minutes respectively. While these may be acceptable in settings where a single diagram needs to be computed offline, we focus our experiments on lower sampling rates to investigate how well these methods perform with relatively few samples.

The sampling strategies themselves are fast in comparison. At $1000$ points this takes less than a second for most methods. The exception is the Voronoi strategy, taking about 1.6 seconds or 7.3 seconds to generate $1000$ samples in local and global variants respectively.
Note that our implementation of the sampling techniques is not optimized, and that samples can be reused in computations that use the same underlying map (only their weights need modification).

\cparagraph{Quality.} 
\begin{figure*}
    % \centering
    \hspace*{\fill}
    \includegraphics{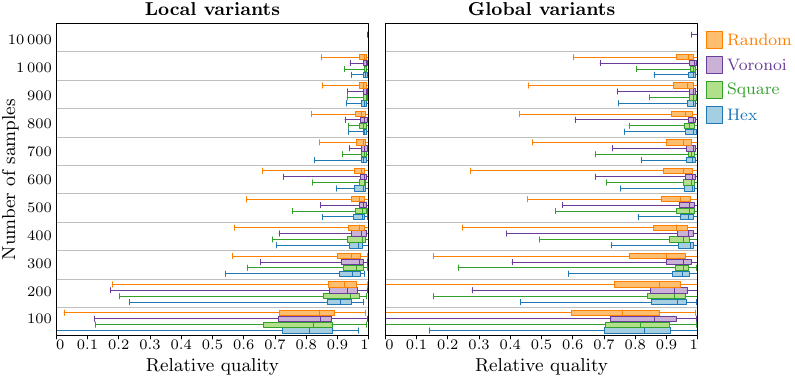}
    \caption{The relative quality obtained by our various sampling strategies. Each box plot summarizes the results over all twenty-four inputs (twelve maps, two classes each to fit a disk to) for all numbers of samples within the range implied by the vertical axis: the indicated number plus/minus 5. The $10\,000$ bin is an exception: it contains the results of one size $10\,000$ Voronoi sample for each input. The Voronoi sampling technique uses 25 iterations.}
    \label{fig:charts}
\end{figure*}
\autoref{fig:charts} shows the sampling strategies and their relative quality for a range of sample counts.
It is readily apparent that the local sampling techniques are more reliable as the results have a smaller spread than the global variants, especially for a larger number of samples.
Not only are they more reliable, the mean quality, too, is generally higher for the local variants.
Indeed, though for 100--400 samples the global Hex technique has a slightly higher median than its local variant, from 500 samples onwards each local technique has a higher median quality.
The Random method benefits most from the point distribution to regions used in the local variants, which is natural as it lessens randomization issues such as large areas without sample points.
However, for the other techniques too, the difference in spread is considerable.
This difference is expected as the point distribution to regions in some sense encodes the region structure and steers samples towards areas where points have possibly different weight.
Indeed, in samples created by local techniques, regions of sufficiently large area are guaranteed to contain a proportionate number of samples, while this is not true for global techniques.
This may cause a global technique to sample little to nothing in a large region important for finding the optimal disk.

The Random technique performs worst of the four and is always outperformed by Voronoi in terms of median quality.
Voronoi and the two grid methods have similar performance; there is no clear best technique among the three.
Hence, from these experiments, using a local grid technique to create a sample of $1000$ points seems a good trade-off between quality and running time. 
In our experiments, the median quality of such a sample was approx.\ $1\%$ lower than the presumed optimal, and the lowest quality was approx.\ $8\%$ lower than the presumed optimal.

\begin{figure*}
    \centering
    \includegraphics[width=0.32\textwidth]{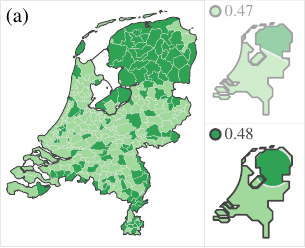}
    \includegraphics[width=0.32\textwidth]{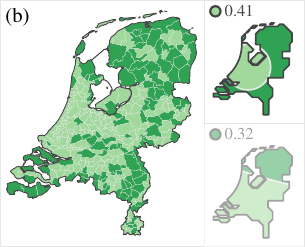}
    \includegraphics[width=0.32\textwidth]{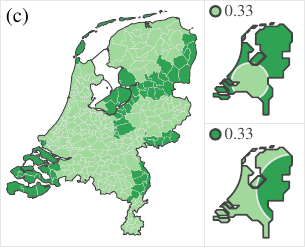}
    
    \vspace{0.5mm}
    
    \includegraphics[width=0.32\textwidth]{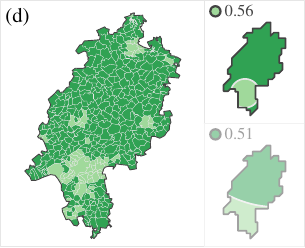}
    \includegraphics[width=0.32\textwidth]{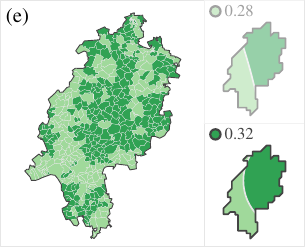}
    \includegraphics[width=0.32\textwidth]{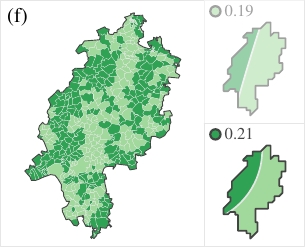}
    \caption{Six choropleths from the experimental evaluation and their corresponding diagrams (one per class that a disk is fit to). Numbers show the normalized score of a disk. Disks computed on a sample of size $1000$ computed using the local Voronoi sampling technique with 25 iterations. Schematic outlines based on results of algorithms by \cite{bmrs2016}.}
    \label{fig:diagrams}
\end{figure*}

\cparagraph{Diagrams.}
\autoref{fig:diagrams} shows chorematic diagrams of six of the twelve choropleths used in the experimental evaluation.
The figure shows two chorematic diagrams per choropleth: one for each class that is to be captured by the disk.
We show the score of the disks after normalization to the range $[-1, 1]$.
For most maps, both disks try to capture the same circular pattern and one of the disks approaches a line.
Map~(c) is an exception: the two disks have near identical scores and capture quite different patterns.
The left four maps---(a), (b), (d), and (e)---have one large roughly circular pattern that the algorithm successfully captures.
Maps (c) and (f) are not easily captured by a single disk, but the algorithm returns reasonable results.
Map~(c) would best be captured by multiple shapes, for example an annulus in the east and a disk in the south-west.
Map~(f) could be captured by, for example, a light green ellipse in the middle.
The choropleth of The Netherlands in \autoref{fig:weights} also comes from the experimental evaluation. The disk shown in that figure has the lowest score (0.19) in our experiments.
Though maps (c) and (f) of \autoref{fig:diagrams} could be captured in a chorematic diagram by using more shapes, the choropleth of \autoref{fig:weights} is a good example of one that does not exhibit any clear pattern that could be captured by few simple shapes.

\section{Conclusion and future work}
\label{sec:conclusion}
We studied the problem of computing chorematic diagrams to represent classed region maps, specifically using disks. After formalizing the problem, we concluded that some form of approximation is necessary in order to end up with an efficient and practical algorithm. 

We approximated the full problem by using various sampling strategies, and evaluated them on a variety of choropleths.
Our experiments show that relatively few samples are necessary to achieve high-quality results.
Additionally, sampling locally in regions, according to the proportion of the map they occupy, results in more reliable and higher quality samples than sampling globally.
Simple grid sampling techniques provide high-quality results, and more sophisticated methods do not seem worth the additional running time they would incur.

\cparagraph{Future work.} 
Our results leave various interesting avenues for future work. For example, one could investigate a hybrid between local and global techniques that samples in the union of all regions with the same value. Another useful direction of research would be to prove formal guarantees on the quality of point samples for solving the kind of shape matching problem we investigate in this paper.

The bottleneck in our pipeline is computing the maximum-weight disk on weighted points.
Investigating how to speed up this step, either by further studying the problem theoretically, or by, for example, investigating whether the algorithm can run effectively on a GPU, is a  possible direction for future work.

Our experiments focus on choropleths with two classes.
Further experiments could investigate how to best address the more general problem mentioned in \autoref{sec:problem} of summarizing choropleths with more than two classes.

As our results also show, actual data is often complex and cannot reasonably be described by just a single disk. Solving our problem for multiple disks, or indeed also other shapes, is an important next step in fully automating the construction of chorematic classed region maps.

Lastly, it would be interesting to compare the diagrams returned by our algorithm with ones created by cartographers to investigate to what extent the cost function we use matches the training and eye of a professional.

\subsection{Acknowledgments}
W. Meulemans is partially supported by the Dutch Research
Council (NWO) under project no. VI.Vidi.223.137. 

\bibliographystyle{apacite}
\bibliography{references_long}

\end{document}